\documentclass[aps,prl,twocolumn,showpacs,preprintnumbers,superscriptaddress,amsmath,amssymb]{revtex4}
\usepackage{graphicx,subfigure}
\usepackage{dcolumn}
\usepackage{bm}
\usepackage{ulem}
\usepackage{color}
\def\be{\begin{equation}}
\def\ee{\end{equation}}
\def\ba{\begin{eqnarray}}
\def\ea{\end{eqnarray}}

\begin{document}

\title{Fermi-surface reconstruction in a smectic phase of a high temperature superconductor}
\author{Hong Yao}
\affiliation{Department of Physics, University of California, Berkeley, CA 94720}
\affiliation{Materials Sciences Division, Lawrence Berkeley National Laboratory, Berkeley, CA 94720}
\author{Dung-Hai Lee}
\affiliation{Department of Physics, University of California, Berkeley, CA 94720}
\affiliation{Materials Sciences Division, Lawrence Berkeley National Laboratory, Berkeley, CA 94720}
\author{Steven Kivelson}
\affiliation{Department of Physics, Stanford University, Stanford, CA 94305}
\date{\today}
\begin{abstract}

It is shown that, in the presence of a moderately strong $C_4$ symmetry  breaking (which could be produced either by lattice orthorhombicity or the presence of an electron nematic phase), a weak, period 4, unidirectional charge density wave (``charge stripe'') order can reconstruct the Fermi surface of a typical hole-doped cuprate  to produce a small electron pocket.  This form of charge density wave order is consistent with that adduced from recent high field NMR experiments in YBCO.
The Fermi pocket
  has an area and effective mass which is a rough
caricature of those seen in recent high field quantum oscillation experiments.


\end{abstract}
\maketitle

There is clear evidence that, in addition to superconductivity, other
broken symmetry electronic phases occur in some circumstances in the cuprate high $T_c$ superconductors. Discovering and characterizing these orders is an integral part of cuprate research.

A high level of excitement has accompanied the discovery\cite{doiron} of quantum oscillations in high quality ``underdoped'' crystals of YBCO
when superconductivity is quenched by high magnetic fields
\cite{leboeuf,sebastian,ramshaw,jaudet,boebinger}.
 What is appealing about these experiments\cite{chakravartycomment}
is that
they potentially reveal the properties of
ordered groundstate phases  which compete with superconductivity.
Specifically, these experiments unambiguously establish the existence of small Fermi pockets with well defined gapless Fermionic quasiparticle excitations. Significantly this ``fermiology'' is quite different from that of  highly overdoped materials as also
 inferred from quantum oscillations\cite{vignolle}.
 Although there
 is
some  uncertainty
 concerning the number of Fermi pockets in underdoped YBCO, there is growing consensus that there is at least one
 ``electron-pocket'' which encloses roughly 2\% of the Brillouin Zone (BZ).
 While there are some
  reasons to worry\cite{boebinger} that this pocket may reflect
 a material specific band which is irrelevant to the physics of the copper-oxide plane, it is widely considered more likely that the electron pocket is the result of a Fermi surface reconstruction brought about by
some form of translation symmetry breaking density wave order.

Theoretical calculations have
 shown that basic features of the above fermiology can be accounted for under the assumption that the system manifests  d-density wave (dDW) order\cite{chakravarty}, spin-spiral order\cite{harrison}, or spin-stripe\cite{kivelson} order\cite{millis}. In contrast, it has been argued\cite{millis} that
 unidirectional CDW (or charge stripe) order cannot produce the requisite
 electron pocket.

In this context, it is significant that a recent high field copper NMR
 experiment\cite{julien} performed on a YBCO sample which does exhibit
 quantum oscillations has lead to the following conclusions:

  1)  There is no effective magnetic field on the Cu sites
   -- this rules out any form of SDW order, although, for special symmetry reasons, it does not rule out dDW order.
  2)
  There is a charge density modulation most probably associated with a site centered unidirectional CDW with period equal to 4 lattice constants \cite{footnote}. Moreover, the amplitude of charge modulation is estimated to be about $0.03\pm 0.01 e$ per planar unit cell.

Very similar forms of charge order have been observed directly\cite{tranquada,abbamonte,hueker,fink,kivelson} with neutron and X-ray scattering for some time in the 214 family of high temperature superconductors, especially ones with generally reduced superconducting transition temperatures.  Such order is often referred to as ``charge stripes.''
 Other studies have revealed a strong enhancement of spin-stripe order in LSCO when superconductivity is suppressed by a moderate magnetic field\cite{lake}. In addition to the above, evidence of a tendency to formation of
charge density wave order with period near 4 lattice constants has been adduced from STM studies of other underdoped cuprates\cite{kapitulnik,yazdani,kohsaka,parker,kivelson},
and of field induced CDW correlations in
vortex cores\cite{hoffman}.
Given these results, it is important to revisit the issue of whether a pure uni-directional CDW order  can produce the requisite electron pocket in YBCO.

 In the following we demonstrate that charge stripes can produce an electron pocket with the indicated characteristics
 so long as, in addition to the CDW order, the electronic structure also breaks 
$C_4$  rotational symmetry sufficiently strongly.
Of course, unidirectional order, itself, breaks this symmetry, as does the orthorhombic structure of YBCO.  However, if these explicit symmetry breaking effects are too weak, as existing calculations suggest, then the strong $C_4$ symmetry breaking could also reflect the existence of an intrinsic electronic tendency to form an ``electron nematic'' state\cite{kfe}, for which there is considerable independent experimental evidence\cite{ando,hinkov,taillefer,kapitulnik,davis,kivelson,vojta}.

 In order to convey our message clearly let us consider the following simple tight-binding Hamiltonian.
 \ba
H&=&\sum_{\vec R,\sigma}\Big\{-t \Big[(1+\phi_N) c_{\vec R,\sigma}^\dagger c_{\vec R+\hat x,\sigma}+(1-\phi_N) c_{\vec R,\sigma}^\dagger c_{\vec R+\hat y,\sigma}\Big]\nonumber\\
&&~~-t^\prime\left[ c_{\vec R,\sigma}^\dagger c_{\vec R+\hat x+\hat y,\sigma}+ c_{\vec R,\sigma}^\dagger c_{\vec R+\hat x-\hat y,\sigma}\right]+h.c.\Big\}\nonumber \\
&-&\sum_{\vec R,\sigma}\Big[ 2V_0\cos(\pi X/2) + 2V_2\cos(\pi X)+\mu\Big] c_{\vec R,\sigma}^\dagger c_{\vec R,\sigma}.
\ea
For $\phi_N=V_0=V_2=0$ the model
is a caricature of the low energy bandstructure  associated with the copper oxide planes
inferred from ARPES studies -- fits to the Fermi surface shape in Bi2201 \cite{bi2212} yield values of $t^\prime/t$ between $-0.2$ and $-0.3$, and in LSCO \cite{lsco} between $-0.3$ and $-0.4$.
Often these fits also invoke a third-neighbor hopping matrix, $t^{\prime\prime}/t$ between 0.1 and 0.2.
For the explicit calculations we have carried out, we have taken $t^\prime/t = -0.3$ or $-0.4$ while $t^{\prime\prime}/t=0$, although within some broad range, the results are not highly sensitive to the choice of band parameters.

 Because YBCO cleaves between highly dissimilar planes, its surface can potentially have a large effect on the surface electronic structure accessible to ARPES studies.
 The analysis of the ARPES spectrum is further complicated by the (possibly substantial) bilayer splitting of the bands expected in YBCO, and the presence in some form of chain-related bands.  Despite all this, most ARPES studies\cite{fujimori,dahm,damasceli} have inferred a rather similar ``normal-state'' Fermi surface structure for YBCO as those seen  in other cuprates with an inferred value of $t'/t \approx -0.4$.  However,  a recent ARPES study\cite{mesot} of YBCO films has observed a Fermi surface shape
 which is rather different from other ARPES results and, correspondingly, from   our assumed model.

\begin{figure}[b]
\subfigure[]{
\includegraphics[scale=0.25]{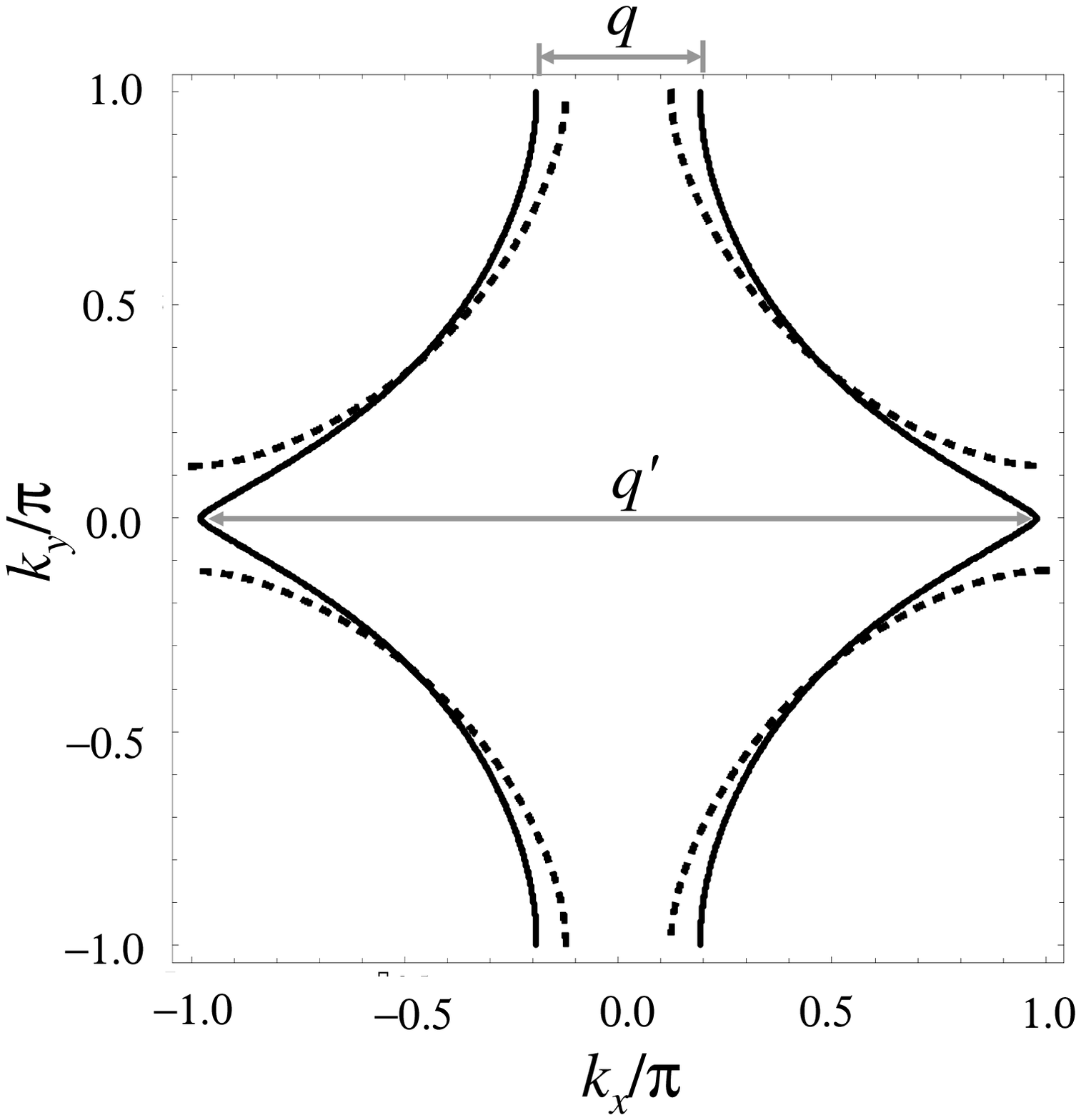}}
\subfigure[]{
\includegraphics[scale=0.25]{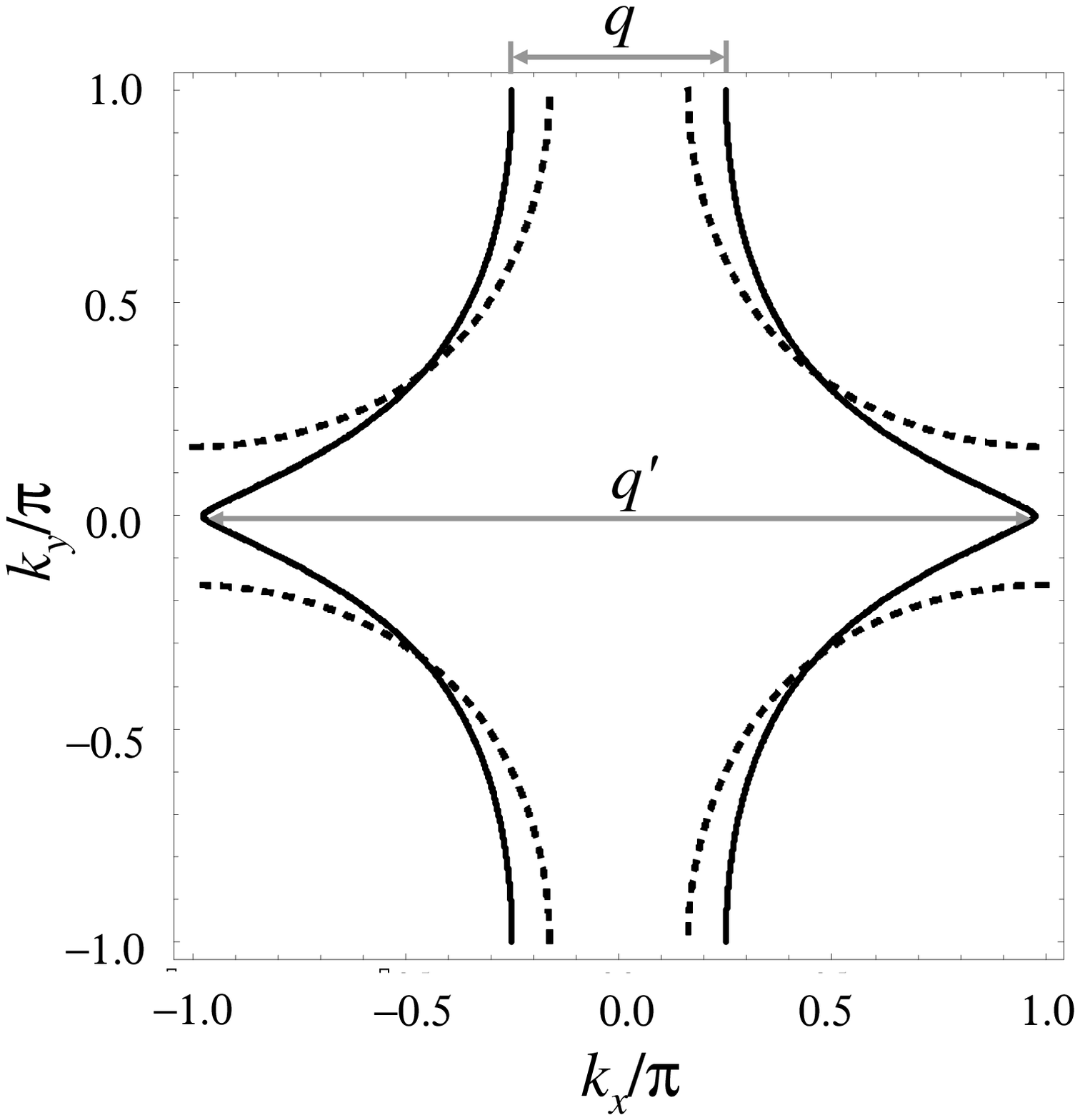}}
\caption{The schematic representation of the Fermi surface with (solid line) and without (dashed line) nematic ordering. In (a), $t^\prime/t=-0.3$ and $\phi_N=0.074$; in (b), $t^\prime/t=-0.4$ and $\phi_N=0.143$ in (a) and (b) respectively. In both cases, the doping is  $x=1/8$.}
\label{fig1}
\end{figure}

The remaining parameters in $H$ represent various broken symmetry.
$\phi_N$ is a measure of the degree of
 asymmetry upon $90^o$ rotation
   in the absence of the density wave order, which reflects both the roughly temperature independent effects of the crystalline anisotropy and a potentially larger and more strongly temperature dependent contribution from any intrinsic tendency toward electron nematic order \cite{fradkin}.
 The term proportional to $V_0$
 is  the fundamental Fourier component  of the effective potential representing translation symmetry breaking corresponding to a period-4 uni-directional CDW;  the choice of the phase of the cosine corresponds  to site-centered charge-stripe order (as indicated by the NMR results).
 The term proportional $V_2$ is generally present as the only allowed harmonic of the fundamental period in a site-centered charge-striped state;  in the special case where we are dealing with the Ortho II phase of YBCO, there is a further contribution to $V_2$ (which is unrelated to density-wave formation) which comes from the existence of
 alternating oxygen-full and oxygen-empty chains in the Ortho II crystal structure. 
 The chemical potential $\mu$ is adjusted so that the
 density of doped holes per planar Cu is
in the neighborhood of 10\%.

\begin{figure}[b]
\subfigure[]{
\includegraphics[scale=0.25]{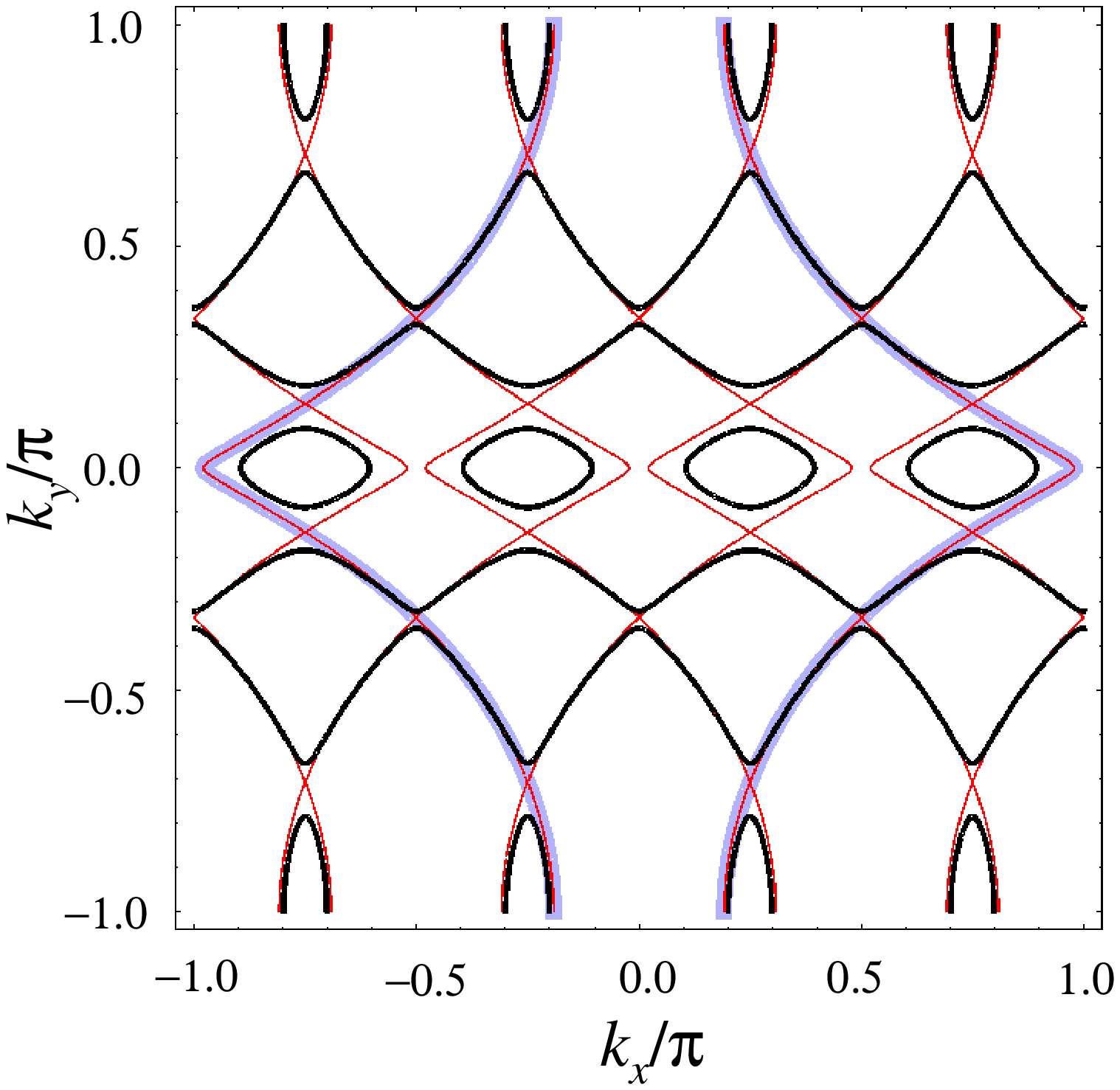}}~~
\subfigure[]{
\includegraphics[scale=0.25]{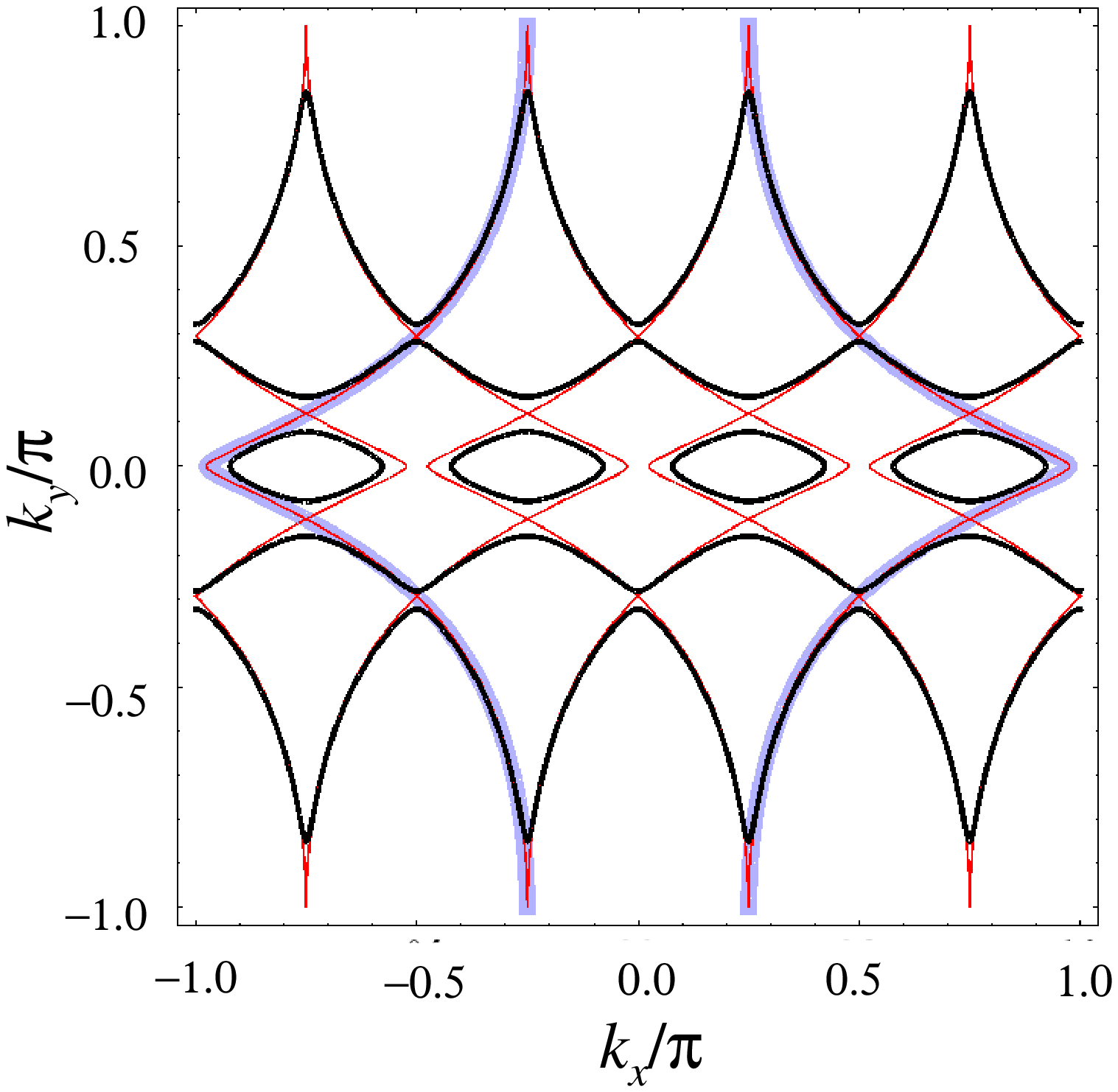}}
\caption{Schematic representation (thick black lines) of the folded Fermi surface with nematic ordering and period-4 charge stripe order and $x=1/8$.  In (a), $t'/t=-0.3$,  $\phi_N=0.074$, $V_0=0.15t$, and $V_2=0.10t$,  while in (b), $t'/t=-0.4$, $\phi_N=0.143$, $V_0=0.11t$ and $V_2=0.09t$, and the electron pocket encloses roughly  1\% of the unreconstructed BZ.  The parameters have been chosen for graphical clarity.  When comparing with experiment we have taken $V_0=0.02t$, and $V_2=0.01t$, so that the electron pockets are roughly 2\% of the BZ, but then the gaps between Fermi surface segments are quite small.
 The thin (red) lines are the zone-folded Fermi surfaces in the limit of vanishing stripe order, $V_0=V_2=0$, and the (blue) shaded line is the unfolded Fermi surface. }
\label{fig2}
\end{figure}

In Fig. \ref{fig1} we show the Fermi surfaces associated with an isotropic normal phase, $\phi_N=V_0=V_2=0$ (dashed line), and with an anisotropic uniform phase,
 $\phi_N\neq 0, V_0=V_2=0$ (solid line) where $\phi_N=0.074$ and $0.143$ in Figs. 1a and 1b, respectively.
Note that we have taken  $\phi_N$
larger than the critical value $\phi_c$ (which are $\phi_c =0.072$ and 0.142 for the
  choice of band parameters in Figs. 1(a) and 1(b), respectively) at which point the Fermi surface
topology changes from closed (and hole-like) to open.
  This sort of Fermi surface reconstruction has been invoked previously\cite{kee} to account (in the context of a weak-coupling RPA treatment) for the nematic order detected\cite{hinkov} in  inelastic neutron scattering studies of somewhat more underdoped YBCO than the material which exhibits quantum oscillations.
Two important
 wave-vectors which characterized the open Fermi surface are
 $q$ and $q^\prime$, the minimum and maximum spanning vectors of the Fermi surfaces in Fig. 1.
 Clearly, $q$ and $q^\prime$ depend
 on the magnitude of $\phi_N$,
 the doping level, and other
  band-structure parameters.

In Figs. 2(a) and 2(b) we show the
  reconstruction
  of the Fermi surface produced by a non-zero CDW potential, $V_{0,2} > 0$.  We
 report the parameters used to construct these
 results in the figure caption.
 However, the qualitative features are easily understood without any calculation,
  by
  folding the Fermi surface according to the period four charge stripe order
  (thin red lines in Fig. 2),
  and reconnecting the Fermi surfaces
  (thick black lines in Fig. 2)
   to produce gaps where ever two folded Fermi surfaces cross.

As a result, so long as $2\pi > q^\prime > 3\pi/2$, a small electron pocket will be formed where the Fermi surface in the presence 
 of nonzero $\phi_N$ 
 makes its closest approach to the $(\pi,0)$ point. The size of this pocket is a decreasing function of $V_0$, and also of $(q^\prime-3\pi/2)$.
 An elelctron pocket with an area around 2\% of the BZ (corresponding to the size observed in the quantum oscillations experiments) can be obtained by choosing small values of $V_0$ and $V_2$;
  for the purposes of computing quantities to compare with experiment, we will take take $V_0=0.02t$ and $V_2=0.01t$
 and the remaining parameters as in Figs (a) and 2(b).  (However, for graphical clarity, in Fig. 2, we have chosen larger values of $V_0$ and $V_2$ to make the character of the Fermi surface reconstruction easier to visualize.)
If $q$ is considerably less than $\pi/2$, then even for non-zero $V_0$ (so long as it is not too large), there will be a hole pocket near the $(0,\pi)$ point in the BZ, as shown in Fig. 2(a). Conversely, if $q$ is close to $\pi/2$, the portion of the Fermi surface near $(0,\pi)$ is nearly nested, so even a small CDW potential ($V_0 \neq 0$) will open a gap in that region of the BZ -- the case shown in Fig. 2(b) \cite{footnote1}.
The remaining portions of Fermi surface are open, as can be seen in the figures, and so cannot contribute to quantum oscillations
(although they can contribute to transport anisotropy).

From the temperature dependence of the quantum oscillations, it is possible to extract the value of the effective mass.  The effective mass that enters the expression for the quantum oscillations is
\be
m^* \equiv \frac{\hbar^{2}}{2\pi}\int  \frac{d k_\parallel}{|\nabla_{\vec k}\epsilon_{\vec k}|}
\ee
where the integral is over the Fermi-surface of the relevant pocket, and $\epsilon_{\vec k}$ is the
dispersion.
With this definition, the electron pockets in Fig. 2(a) and 2(b) have effective mass $m^*/m_e=0.19 eV/t$ and $0.22eV/t$, respectively, where $m_e$ is the bare electron mass.  The measured effective mass in the quantum oscillations is typically about twice the free electron mass, although it varies somewhat with doping concentration.
 This would agree
with the above result if we set
$t\approx 0.1 eV$
\cite{footnote2}.

The same effective mass enters the contribution of the electron pocket to the specific heat.
As we are ignoring all effects of  bilayer coupling, there
are two identical electron pockets  per Cu-O
bilayer  in the reduced BZ ($-\pi < k_y \leq \pi$ and $-\pi/4 < k_x \leq \pi/4$).
At low temperatures, this gives rise to an electron pocket contribution to the specific heat, $C_{el-poc} = \gamma_{el-poc} T$ with $\gamma_{el-poc} =  2.92 (m^\ast/m_e)$
mJ$\cdot \textrm{K}^{-2}\cdot $mol$^{-1}$, which is comparable to the total specific heat measured\cite{boebinger} in the relevant field range - for instance\cite{boebinger}, at 40T, $\gamma_{exp} \approx 4.2$ mJ$\cdot \textrm{K}^{-2}\cdot $mol$^{-1}$. However, the total contribution to the specific heat from all the portions of the Fermi surface in the reconstructed band-structure is $\gamma_{th}=3.28(eV/t)$mJ$\cdot \textrm{K}^{-2}\cdot $mol$^{-1}$ for Fig. 2(a) and $4.03(eV/t)$mJ$\cdot \textrm{K}^{-2}\cdot $mol$^{-1}$ for Fig. 2(b), which for any reasonable value of $t$ are considerably larger than the measured specific heat.  This is a serious problem with interpreting experiments in YBCO in terms of the simple model considered here.  Possibly, as conjectured by the authors of Ref. \cite{boebinger}, superconducting fluctuations still produce
 a residual gap on at least some portions of the ``normal state'' Fermi surface.

The amplitude of the charge modulation, defined as the difference between the highest and lowest site charge density, is calculated to be about 0.05$e$ in both cases of Fig. 2(a) and Fig. 2(b), which is reasonably close to the value 0.03$\pm 0.01 e$, adduced\cite{julien} from the high field NMR experiments. The relative modulation of density of states at the Fermi level ($\rho$) is 
computed to be about $\delta\rho/\bar \rho \approx 0.1$ for Fig. 2(a) and $\delta\rho/\bar \rho \approx 0.04$ for Fig. 2(b), from which the corresponding variation of the Knight shift can be predicted. (The spatial average of $\rho$ is $\bar \rho\approx 0.5/t$.)

Obviously there are many further effects that have not been taken into account in the present study.
For example we have ignored the effects of fluctuations, and of incipient spin order.  Moreover, there is some evidence\cite{reznik}, from neutron scattering studies of phonon anomalies in YBCO, that there is a strong tendency toward $2k_F$ charge density wave formation along the stripe direction ({\it i.e.} along the $y$ direction), which we have neglected as well.  Certainly, the rough consistency we have demonstrated with experiment is far from a unique property of our simple model. For example, at least as good agreement with the quantum oscillation data has been obtained under the assumption of dDW order\cite{chakravarty}.

In addition, there are many aspects of the this problem that are perplexing, and may well imply that much more subtle effects are at play.  A few of the more vexing are:  1) It is not at all clear how to reconcile the existence of an electron pocket in the antinodal region, near $(\pi,0)$, with clear spectroscopic evidence of a well developed pseudo-gap in this region.\cite{hehe}  The NMR data suggests that charge stripe order condenses only when the field exceeds a large critical value, of order 20T, and possibly this field is large enough to quench the antinodal pseudogap.
 However, an intuitive link has been suggested between the existence of an electron pocket and the observed sign change of the Hall resistance and thermopower\cite{oldtaillefer, newtaillefer} below a critical temperature, $T_0$. This sign change can be observed at relatively low fields in YBCO with $x$ near 1/8, where spectroscopic studies still show a well developed pseudo-gap\cite{valla,he,basov}. Moreover, similar signatures are seen in the Hall and thermopower\cite{basov,oldtaillefer, newtaillefer} in Eu doped LSCO and LBCO, where the superconducting $T_c$ is suppressed  and charge stripe order is stabilized by the LTT crystal structure, even in zero field and where, moreover, the existence of an antinodal pseudogap is well established\cite{basov}.
 2)  It is observed that at lower doping, $x \approx 8$\%, YBCO does not exhibit quantum oscillations.  It has been suggested\cite{proust} that this is due to the electron-pocket vanishing by a Lifshitz transition.  While such a transition is possible in the present scenario, it does not seem natural.  However, at this low doping, spin order is observed\cite{sonier,sanna,haug}, which at short distances can be thought of as spin-stripe order, but at long distances is characterized as a spin-glass.  Possibly the scattering of electron quasiparticles from the spin-glass order or slow fluctuations quenches the quantum oscillations.
 3)  Most importantly, the specific heat data\cite{boebinger} strongly suggest that substantial local superconducting order, and the corresponding gapping of the quasiparticle spectrum, survives even at the high fields where quantum oscillations are observed.  How any of the physics we have addressed plays out in the presence of strong superconducting fluctuations is unclear to us at present.
 4)  Finally, it is important to stress that the model we have treated omits many material specific details, such as effects of bilayer splittings and the presence of chain bands, which even if they are not conceptually important, can greatly affect the outcomes of specific experiments.

 \acknowledgements{We thank M. -H. Julien for allowing us to see his data prior to publicaiton, and M.-H. Julien, C. Proust, J. Tranquada, E. Fradkin, and R.-H. He for insightful discussions. DHL and HY were supported by DOE grant number DE-AC02-05CH11231 at UCB and  SK was supported, in part, by DE-FG02-06ER46287.}


\begin{thebibliography}{22}
\bibitem{footnote} There is another form of CDW order that is consistent with the NMR data which has ordering vector $(\pi/2,\pi)$,
in which there is a charge modulation with period 2 along the stripe direction on top of an underlying period 4 site-centered stripe state.
\bibitem{footnote1} A third, qualitatively distinct situation arises  when $q$ is considerably larger than $\pi/2$, but this does not appear to be pertinent to the cuprates, so it will not be discussed here.
\end{thebibliography}
\end{document}